\documentclass[12pt,groupedaddress,showpacsaps,amsfonts,amssymb,floatfix]{iopart}
\usepackage{graphicx}
\usepackage{dcolumn}
\usepackage{bm}
\usepackage{epsfig}
\usepackage{epstopdf}\usepackage{float}
\usepackage{color} 
\newcommand{\qvec}{{\bf q}}

\begin{document} 
\title{Gutzwiller Charge Phase Diagram of Cuprates, including Electron-Phonon Coupling Effects}
\author{R.S. Markiewicz}
\address{Physics Department, Northeastern University, Boston MA 02115, USA}
\author{G. Seibold}
\address{Institut F\"ur Physik, BTU Cottbus-Senftenberg, PBox 101344, 03013 Cottbus,
Germany}   
\author{J. Lorenzana}
\address{ISC-CNR and Dipartimento di Fisica, Universit\`a di Roma
``La Sapienza'', P. Aldo Moro 2, 00185 Roma, Italy}
\author{A. Bansil}
\address{Physics Department, Northeastern University, Boston MA 02115, USA}
\begin{abstract}
Besides significant electronic correlations, high-temperature superconductors
also show a strong coupling of electrons to a number of lattice modes.
Combined with the experimental detection of electronic inhomogeneities 
and ordering phenomena in many high-T$_c$ compounds,  
these features raise the question as to what extent phonons are 
involved in the associated instabilities. Here we address this problem
based on the Hubbard model including a coupling to phonons in order to 
capture several salient features of the phase diagram of hole-doped cuprates.
Charge degrees of freedom, which are suppressed by the large Hubbard $U$ near
half-filling,  are found to become active at a fairly low doping level. 
We find that possible charge order is mainly driven by Fermi surface 
nesting, with competition between a near-$(\pi,
\pi)$ order at low doping and antinodal nesting at higher doping, 
very similar to the momentum structure
of magnetic fluctuations. The resulting nesting vectors are generally  
consistent with photoemission and tunneling observations, 
evidence for charge density wave (CDW) order in YBa$_2$Cu$_3$O$_{7- \delta}$ including Kohn 
anomalies, and suggestions of competition between one- and two-$q$-vector 
nesting.
\end{abstract} 
\maketitle
\setcounter{footnote}{0}

\section{Introduction}
The existence of charge density wave (CDW) order 
is now well established for a large class of high-temperature superconductor
materials. Starting from the pioneering studies in lanthanum cuprates
\cite{Tranq}, recent resonant and hard x-ray
scattering data have revealed CDW modulations
also in YBCO \cite{YCDW2,YCDW3,YCDW4,YCDW5,YCDW6} and Bi2201 \cite{comin} compounds. However, whereas in lanthanum cuprates a concomitant spin scattering 
with twice the period of the CDW suggests the formation of charge-spin
stripes \cite{LoSe}, there seems to be no apparent relationship between the two
periodicities in the YBCO and Bi2201 materials. Instead, the analysis of 
resonant x-ray scattering and angle-resolved photoemission spectra indicates that the nesting properties of the underlying Fermi surface (FS) or 
`Fermi arcs' are at play.
This has led to proposals\cite{wang14,allais14} in which 
CDW formation related to FS features is driven by
magnetic interactions.  
In this paper we examine the simpler possibility that electron-phonon
interactions could play a role in the formation of the observed CDW
modulations.  Earlier proposals of phonon induced CDWs\cite{cas95}
did not rely on nesting features.

We adopt an intermediate coupling approach in our analysis in this study.\cite{AIP} This is justified by recent
studies\cite{correl1,Gzm2,tanmoyop,ASWT,correl2,correl3}, which have
indicated that correlations in the cuprates are not as strong as initially
believed, and that cuprates fall, instead, in an intermediate coupling regime, with $6\le U/t \le 9$, where   
$U$ is the effective Hubbard interaction and $t$ is the nearest-neighbor hopping parameter. We have shown \cite{AIP} that
intermediate coupling corresponds approximately to $4\le U/t \le 13.6=U_{BR}/t$, where $U_{BR}$ is the mean-field
Brinkman-Rice energy where double occupancy vanishes.\cite{br,Gzm2}  
In this regime competing phase transitions often evolve from Stoner instabilities, which can be described by Hartree-Fock (HF) or, better,
Gutzwiller approximation (GA) based calculations.\cite{footRef,TBPS} 
Furthermore, at large doping cuprates behave as
Fermi liquids, so that one can hope to obtain information on
ordered phases by studying the instabilities that disrupt the Fermi
liquid behavior, provided the correlated physics is included in the
analysis. We have shown that in the weak and intermediate coupling range,  peaks in the bare susceptibility of 2D materials, which determine magnetic instabilities, form a map of the FS, and 
the dominant instabilities are generally related to the {\it double nesting} 
features\cite{Gzm2,Gzm1}, where two branches of the map cross.  In particular, the $T=0$ magnetic phase 
diagram of the cuprates was derived using the time-dependent GA 
(TDGA).\cite{Gzm1} In the electron-doped cuprates, the magnetic phase 
diagram is dominated at all dopings by a commensurate $(\pi ,\pi )$ antiferromagnetic (AFM) order.\cite{Gzm1} In contrast, for 
the hole doped cuprates, we find a wide doping range over which the magnetic order is incommensurate.

Given the large onsite Coulomb repulsion (Hubbard $U$), magnetic order should be favored near half-filling, but for larger hole doping the experimental evidence is more consistent with incommensurate charge order as noted above.
In this paper, we apply the TDGA technique to examine the charge response, 
including effects of finite electron-phonon coupling of  
Su-Schrieffer-Heeger form\cite{SSH}. When phonons are included, we find charge density wave (CDW) phases with nesting vectors similar to those for the magnetic instabilities, which are controlled by a generalized Stoner criterion and the double nesting features in the susceptibility. We present the full evolution with doping of the leading charge-phonon instabilities for several families of cuprates, including La$_{2-x}$A$_x$CuO$_{4+\delta}$, A = Sr 
(LSCO) or Ba (LBCO) and Bi$_2$Sr$_2$CuO$_6$ (Bi2201).

The magnetic [charge] instabilities are usually determined by zeroes of the 
Stoner denominator,\cite{Gzm1}   
\begin{equation}
1-[+]U_{\rm eff}(q) \chi_0(q,\omega =0).
\label{eq:G1}
\end{equation}
In a HF plus random-phase approximation (RPA) calculation
$U_{\rm eff}(q)$ would simply be the Hubbard $U$ and $\chi_0(q,\omega =0)$
the susceptibility for local magnetic [charge] fluctuations. But in the TDGA the situation is more complex since local and transitive fluctuations are coupled so that $U_{\rm eff}(q)$ depends on the corresponding susceptibilities and the associated coupling constants. Here, by ‘local’ we mean that the phonon modulates the on-site energies, as in the Holstein model, while ‘transitive’ means that the hopping parameters are modified, as in the Su-Schrieffer-Heeger model.

Thus, the leading HF+RPA instability is simply associated with the maximum of the bare susceptibility $\chi_{0M}={\rm max}_q\, \chi_0(q,0)$, while the leading 
Gutzwiller instability can be shifted by the $q$-dependence of $U_{\rm GA}(q)$.  
It is clear that such instabilites cannot arise for a purely local electron-phonon coupling since local charge fluctuations are significantly suppressed in the presence of correlations. However, the situation is different
for the coupling of phonons to transitive fluctuations,\cite{odlsg} 
which can induce a CDW in a system with sizeable electronic correlations for moderate values of the electron-phonon coupling.

This paper is organized as follows.  Section \ref{secform} 
describes our model system, and focuses on the
electron-phonon coupling related aspects. 
Section \ref{secres} presents results for the renormalized phonon dispersions
and the resulting charge phase 
diagrams for various types of high-$T_c$ materials.  
In Section \ref{secexp} we compare our results with experiments on
the cuprates, in particular we examine evidence for a crossover between 
charge and magnetic instabilities in the underdoped compounds, 
the doping dependence of nesting vectors, and their relationship to Kohn anomalies and soft phonons. We conclude our discussion in Section \ref{secconcl}. In \ref{subsecga} we describe our TDGA formalism for CDWs, and in \ref{appb} we discuss one- vs two-$q$ nesting. 
Further applications of the model are briefly considered in the Supplementary Online Materials (SOM), and include extensions to photoemission and tunneling studies [Section SOM1], a search for purely electronic CDWs  [Section SOM2], and a discussion on stacks of Kohn anomalies [Section SOM3].

\section{Model and Formalism}\label{secform}

Our investigations are based on the following Hamiltonian
\begin{equation}\label{eq:mod}
H = H_{e} + H_{el-ph} + H_{ph}
\end{equation}
where $H_{e}$ denotes the Hubbard model, $H_{e-ph}$ the coupling
between electrons and phonons and $H_{ph}$ is the bare phonon part.
In the Hubbard model
\begin{displaymath}                                  
H_{e}=\sum_{ij,\sigma}t_{ij} c_{i,\sigma}^\dagger c_{j,\sigma}  
+  U\sum_i n_{i,\uparrow}n_{i,\downarrow}                       
\end{displaymath}                                                  
$c_{i,\sigma}^{(\dagger)}$ destroys (creates) an electron on lattice site $R_i$ and $n_{i,\sigma}=c_{i,\sigma}^\dagger c_{i,\sigma}$. We incorporate band structure effects as in our earlier magnetic phase diagram calculations\cite{Gzm1},
by using for the hopping parameters $t_{ij}$ a one-band tight-binding fit 
to the local density approximation (LDA) dispersion\cite{Arun3}  for the 
single-layer cuprates LSCO, Nd$_{2-x}$Ce$_x$CuO$_4$ (NCCO), and Bi2201.  
For convenience, the band parameters are listed in Table I, and the dispersion is given by
\begin{eqnarray}
     E({\bf k})=-2t[c_x(a)+c_y(a)]-4t'c_x(a)c_y(a) \nonumber \\
     -2t''[c_x(2a)+c_y(2a)] \nonumber \\
     -4t'''[c_x(2a)c_y(a)+c_y(2a)c_x(a)] \> ,
\label{eq:0a}
\end{eqnarray}
where
\begin{equation}
     c_i(\alpha a)\equiv \cos(\alpha k_ia) \>,
\label{eq:0b}
\end{equation}
and $\alpha$ is an integer.  Here $k_z$ dispersion is neglected, 
approximating the cuprates as 2D.\cite{AB1a,AB1b} Interaction effects are incorporated via the TDGA, which is used for deriving the charge susceptibility. Details of the formalism are discussed in Refs. \cite{odlsg,DLGS,SeiLo}, and are also summarized in \ref{subsecga}.

\begin{table}
\caption{I. Band Parameter Sets}

\begin{tabular}{||c|c|c|c||}
\hline
Parameter           &       NCCO       &     Bi2201   &  LSCO   
   \\      \hline          
t        &    420~meV&   435~meV   & 419.5~meV \\  \hline  
t'       &   -100    &  -120       & -37.5    \\  \hline
t''      &     65    &    40       &  18      \\  \hline
t'''     &      7.5  &     0       &  34      \\  \hline
\hline
\end{tabular}  
\end{table}

It is obvious that phonons which couple to the
local charge density have only a negligible effect on the associated electronic
fluctuations which are strongly suppressed by the onsite correlation $U$.
The situation is different for phonons which couple to transitive
fluctuations as has been shown in Ref. \cite{odlsg}. For this reason, the electron-phonon coupling $H_{el-ph}$ in Eq.~\ref{eq:mod} is described by a generic Su-Schrieffer-Heeger\cite{SSH} phonon model
consisting of only longitudinal and [in-plane] transverse acoustic branches, 
ionic mass $M$, and electron-phonon coupling, which arises through a modulation of the hopping integral $\delta t$. The key ingredient is that the phonons modulate the hopping parameters with `longitudinal' and ‘transverse’ modulations. Here, longitudinal refers to modulation $\delta {\bf a}$ along the phonon propagation direction, and transverse to those at right angles to the propagation direction. The corresponding operator is:
\begin{equation}
H_{el-ph}=-\sum_{ij} \frac{t_{ij}\gamma_{ij}}{r_{ij}}  \sum_{\sigma \mu=x,y} (u^\mu_{j}-u^\mu_i)
(c_{i\sigma}^\dagger c_{j\sigma} + h.c.) 
\label{eq:SSH}
\end{equation}
where $u^\mu_{i}$ denotes the displacement of the atom at site $R_i$ in direction $\mu$, and 
\begin{equation}
\delta t^{\mu}_{ij}/t_{ij}=-\gamma_{ij}\delta r_{ij}/r_{ij}, 
\label{eq:SSHg}
\end{equation}
with $\delta r_{ij}=(u^\mu_{j}-u^\mu_i)$, and the distance between atoms $i$ and $j$ is $r_{ij}=R_i-R_j$. Finally, the phonon part is given by
\begin{equation}\label{eq:phonly}
H_{ph}= \frac1{2N}\sum_{\alpha\beta\qvec} u^\alpha_{\qvec} K_{\alpha\beta\qvec} 
u^\beta_{-\qvec}
+ \frac1{2N}\sum_{\alpha\qvec} p^\alpha_{\qvec} \frac{1}{M} p^\alpha_{-\qvec}
\end{equation}
which can be diagonalized to yield the bare phonon frequencies $[\Omega_{q\mu}^0]^2=2(K/M)_{\mu}(2-\cos(q_xa)-\cos(q_ya))$.
Here $K_{\mu}$ and $M_{\mu}$ denote the effective spring constant and ionic mass for the longitudinal and in-plane transverse ($\mu$ = L [T]) acoustic mode, respectively. Since the hopping parameters are labeled $t$ for the nearest-neighbor, $t'$ for the second-nearest-neighbor, etc., we label the corresponding $\gamma$'s as $\gamma$, $\gamma'$, and so on.  The nearest-neighbor electron-phonon coupling constant thus becomes $\lambda_{ep}=4\bar N(0)g^2/K$,\cite{footlep,MCA1,MCA2,footMcM,McM} with average density-of-states $\bar N(0)\sim 2/8t$, and electron-phonon coupling $g=\gamma t/a$, or
\begin{equation}
\lambda_{ep}=\frac{\gamma^2t}{Ka^2}.
\label{eq:29c}
\end{equation}
Note that $\lambda_{ep}$ is independent of the ionic mass.

In reality, a strong modulation of $\delta t$ can be produced by several phonons, including those involving motion of oxygen atoms perpendicular to the CuO$_2$ planes\cite{SeGL,gml11}.  Hence, we assume the same bare acoustic frequencies for all cuprates, adjusted to approximate oxygen modes in undoped La$_2$CuO$_4$ [as a generic single-layer cuprate],\cite{WYK} which gives $(K/M)_{LA}=2(K/M)_{TA}\equiv [\Omega_L^{0}]^2$, taking $\Omega_L^0=12.4$~meV, and $M$ is the oxygen mass (where LA[TA] = longitudinal [transverse] acoustic mode). For these parameters the bare dispersion is shown in Figs.~\ref{fig:5} and \ref{fig:6} (dashed lines) for a selected cut through the Brillouin zone.

Thus, the model is completely specified in terms of the coupling constant $\lambda_{ep}$, Eq.~\ref{eq:29c}, which in turn is known up to the hopping coefficient $\gamma$ (where $t\sim 1/r^{\gamma}$). For Bi2201, with $t=435$~meV, 
$\lambda_{ep0}=\lambda_{ep}/\gamma^2=0.047$.  Although the $\gamma$s have proven difficult to calculate,\cite{Alig,NKF}  in the large distance limit, direct wave function overlap on different atoms falls off exponentially with $r$, and in the cuprates, all hopping parameters except the closest $Cu-O$ hopping $t_{CuO}$ are dominated by an indirect chain of hoppings,\cite{OKA2}  yielding for the nearest-neighbor [Cu-Cu] hopping $t\simeq t_{CuO}^2/\Delta$, where $\Delta$ is the on-site energy difference between Cu and O.  Now $t_{CuO}\sim r^{-\gamma_{CuO}}$, with $\gamma_{CuO}\simeq 3-3.5.$\cite{Alig,NKF}  The problem is, what is $\Delta$, and how does it vary with $r$?  We note the following: (1) $\Delta$ remains finite for infinite separation, suggesting a weak $r$ dependence; (2) In a strongly correlated system, the Cu on-site energy has a contribution from $U$.  The Cu$-d$ electrons in the anti-bonding CuO$_2$ band are mostly second electrons on each site, so that $\Delta$ would be dominated by the $U$-term;  (3) We expect $U$ to decrease with decreasing $r$, due to enhanced screening effects.  Thus, $\Delta$ should probably {\it decrease} as $r$ decreases, suggesting $\gamma>6-7$.  Thus, we estimate that $\gamma$ lies in the range between $\gamma_{CuO}\sim$ 3.25 and 10.  In the present paper, we assume $\gamma'=\gamma''=...=0$, unless noted otherwise.

From Eq.~\ref{eq:29c} we estimate that $\lambda_{ep}$ lies between $0.57$ (using $\gamma=3.25$) and 5.4 (for $\gamma=10$).  
Note that the lower estimate is comparable to values in the literature, assuming a modest anisotropy. Recent linear response calculations have found Brillouin zone-averaged coupling strengths $\lambda_{ave}$ of $\sim
0.4$ for optimally doped LSCO\cite{Giustino} and
Ca$_{0.27}$Sr$_{0.63}$CuO$_2$,\cite{Lep1} and 0.27 for
YBa$_2$Cu$_3$O$_7$.\cite{Lep2,Lep3}  However, $\lambda$ has a strong
momentum dependence, for instance, the nodal value is
considerably smaller: in LSCO, $\lambda_{nodal}$ = 0.14 - 0.22 at
optimal doping, and 0.14 - 0.20 in the overdoped regime.\cite{Giustino}
Since nodal electrons dominate transport\cite{tanmoyop},
this accounts for the smaller $\lambda$’s estimated from transport measurements.  Further, several calculations suggest that correlation effects can enhance the anisotropy of electron-phonon coupling, generally leading to a larger value for AN nesting\cite{DLGS,RG1,trans1,QMC1}.  Finally, using the correct density-of-states rather than the average $\bar N(0)$ will further increase $\lambda_{ep}$ in the doping regime near the Van Hove singularity.
Thus our lower estimate for $\lambda_{ANN}$ is likely to be on the conservative side. The electron-phonon interaction renormalizes the frequencies of
bare phonons as well as the charge response of the electrons.
In the presence of intermediate electronic correlations, both
the phonon propagator as well as the charge susceptibility
can be conveniently evaluated via the time-dependent
Gutzwiller approximation (TDGA) as outlined in \ref{subsecga}.

\section{Results}\label{secres}
\subsection{Renormalized phonon dispersions}\label{secres1}

The TDGA charge correlations induce a screening of the phonons and
thus renormalize the bare dispersion $\Omega_{0q\mu}$
according to  
\begin{equation}
\Omega_{q\mu}^{2}=\Omega_{0q\mu}^{2}+\frac{\delta K_{\mu\mu}}{M} .
\label{eq:18}
\end{equation}
A detailed derivation of the correction to the elastic
spring constant $\delta K_{\mu\mu}$
within the TDGA is given in \ref{subsecga} (cf. also Ref. \cite{gml11}).

There is a close connection between the present results and the earlier magnetic results in 
that the instability is controlled by an effective Stoner criterion.  To see this, we write
\begin{equation}
\Omega_{q}^2=\Omega_{0q}^{2}[1-U_{eff,q}\chi_{0q}].
\label{eq:26}
\end{equation}
Then, if for some $q$, $U_{eff,q}\chi_{0q}=1$, the corresponding phonon frequency will vanish, leading to an instability.

\begin{figure}
\includegraphics[width=9cm,clip=true]{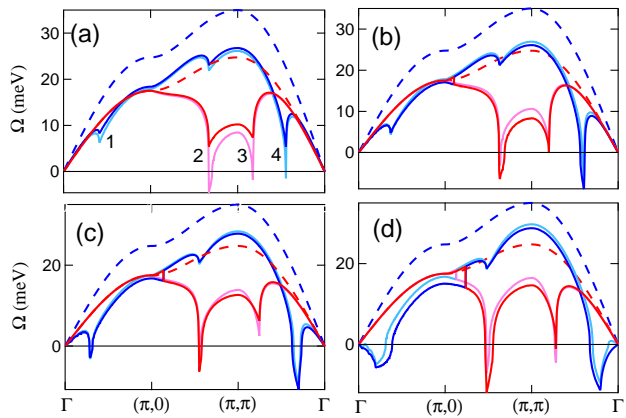}
\caption{(Color online.)
Bare phonon dispersion (dashed line) compared to dressed dispersion assuming $U/U_{BR}$
= 0.20 (light lines) or 0.60 (dark lines) at a series of hole dopings $x$ = (a) 0.05, (b) 0.10, (c) 0.20, (d) 0.30.  Longitudinal [transverse] phonons are shown in shades of blue [red].  Material parameters used are appropriate for Bi-2201, for which $U_{BR}=13.6t$. Only modulation of the nearest-neighbor hopping $t$ is included, with doping independent $\gamma =3.25$. Real $\Omega_{ph}$'s are
plotted as positive numbers, imaginary $\Omega_{ph}$'s as negative.
}
\label{fig:5}
\end{figure}
Figure~\ref{fig:5} compares the bare and renormalized LA and [in-plane] TA phonon
frequencies in Bi2201.  Although the modes along the $(\pi ,0)\rightarrow (\pi ,\pi )$-branch are mixed, these are labeled as being predominantly longitudinal or transverse.  The sharp dips in the dressed frequencies are
caused by peaks in the bare susceptibility associated with FS nesting.  
Each peak in $\chi_{0{\bf q}}$
leads to a prominent Kohn anomaly in the phonon spectrum, which can lead to an instability
if the renormalized $\Omega_{q\mu}^2$ becomes negative. By comparing 
the present results with
earlier calculations for magnetic stripes,\cite{Gzm1} we find that the instabilities fall at
nearly the same $q$-values for both kinds of stripes as a function of doping, being
controlled by the same FS nesting.
However, the relative strengths of the instabilities can be modulated by the electron-phonon coupling. Additionally, the detailed analysis of the
instabilities in the charge sector 
reveals that they  tend to favor lower-q values due to the
momentum structure of $\Omega_{0q\mu}$.

Thus, in the magnetic phase diagram an instability of the Fermi liquid
towards vertical incommensurate order near $(\pi ,\pi )$ 
[$(\pi -\delta ,\pi )$]  was found at low hole doping, $x\le 0.16$,
while an instability towards a diagonal incommensurate phase near $\Gamma$ [$(\delta ,\delta )$] was found from $x=0.16$ to $x=x_{VHS}=0.42$, where $x_{VHS}$ is the doping where the Fermi level crosses the Van Hove singularity (VHS).  The latter instability is associated with nesting of the flat FS sections in the antinodal parts of the FS, and hence is referred to as antinodal nesting (ANN). In contrast, we will refer to the former phase as near nodal nesting (NNN).  Both of these instabilities are controlled by double nesting (simultaneous nesting of two sections of FS at the same $q$). We find that the same instabilities dominate the charge phase diagram, but that additional single nesting charge instabilities start to become competitive because they correspond to lower $\Omega_{0q\mu}$. Near $(\pi ,\pi )$ there are both double nesting vertical [$(\pi -\delta ,\pi )$] (feature 2 in Fig.~\ref{fig:5}(a)) and single nesting diagonal [$(\pi -\delta,\pi -\delta )$] instabilities (feature 3).  Likewise, the ANN instabilities can be either double nesting diagonal [$(\delta ,\delta )$] (feature 4) or single nesting vertical [$(\delta ,0)$] (feature 1). Proximity to the $\Gamma$-point tends to favor ANN nesting over NNN, particularly in Bi2201, where $t'$ is larger and the $(\pi,\pi)$-plateau is weaker.  Thus in Bi2201 the near-$(\pi ,\pi )$ Kohn anomalies are always subdominant, except for small $U$ at extremely low doping.  On the other hand, the diagonal charge ANN anomaly is unstable over the full doping range $x\le 0.4$.  While for the present choice of electron-phonon couplings the diagonal instability is dominant, for other choices the vertical ANN instability wins out at moderate doping ($x<0.2$). Note in Fig.~\ref{fig:5} that the $q$ vector of the leading instability shifts toward $\Gamma$ as doping increases toward $x_{VHS}$, as found previously for magnetic instabilities.  While there are Kohn anomalies in both phonon branches, for all dopings, the ANN instabilities are in the LA phonon branch.

\begin{figure}
\leavevmode  
\epsfig{file=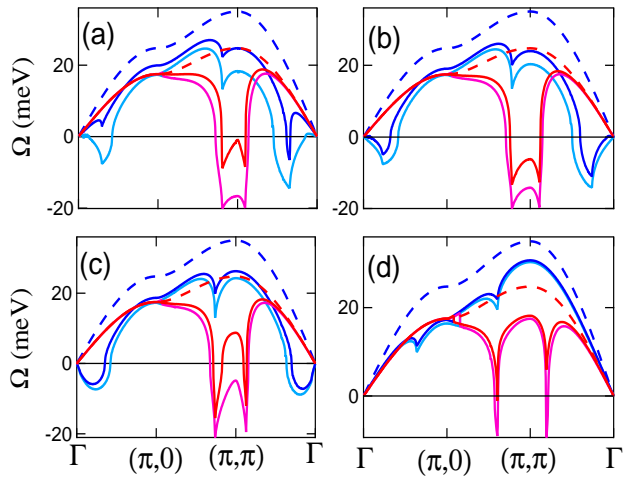,height=7cm,width=9cm, angle=0}
\vskip0.5cm
\caption{(Color online.)
Bare phonon dispersion (dashed line) compared to dressed dispersion assuming $U/U_{BR}$ = 0.20 (light lines) or 0.50 (dark lines) at a series of hole dopings $x$ = (a) 0.05, (b) 0.10, (c) 0.20, (d) 0.40.  Longitudinal [transverse] phonons in shades of blue [red]. Material parameters appropriate for LSCO, assuming $\gamma'=\gamma$.}
\label{fig:6}
\end{figure}

Fig.~\ref{fig:6} presents phonon dispersion maps for LSCO. Unlike the magnetic phase diagram, LSCO behaves similarly to the other cuprates, despite the smaller value of $t'$.  With doping, there is a competition between a predominantly TA phonon soft mode near $(\pi ,\pi )$ and an LA soft mode near $\Gamma$.  For low ($x<0.05$) or very high ($x\ge 0.24$) doping the vertical TA instability at $(\pi ,\pi-\delta )$ is dominant, while at intermediate $x$ the diagonal 
LA instability closer to $\Gamma$ dominates.

\subsection{Charge phase diagrams}\label{secres2}

Figs.~\ref{fig:5} and \ref{fig:6} show that there is competition between Mott physics and strong electron-phonon coupling.  
At each doping and fixed electron-phonon coupling $\lambda_{ep}$ 
there is a critical $U_c(x)$ such that the charge order (CO) phase exists only for values of $U<U_c$.  [For the cuprates, $U\sim 0.6U_{BR}$.]  As $U\rightarrow U_c$, generally the $q$ at the instability is 
sharply defined, and mainly determined by peaks in the bare susceptibility
$\tilde\chi_{0{\bf q}}$. Therefore, the corresponding momenta
match the nesting curves introduced in Ref.~\cite{Gzm1} for 
magnetic instabilities, for which analytic formulas are available.

\begin{figure}
\leavevmode  
\centering
\resizebox{8.8cm}{!}{\includegraphics{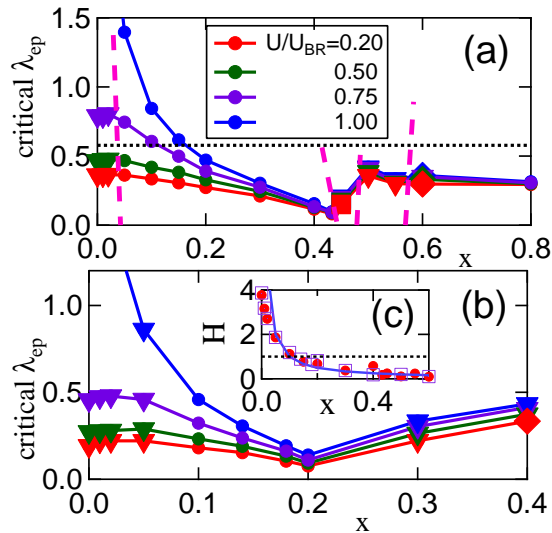}}
\vskip0.5cm
\caption{(Color online.)
Phase diagrams of the critical electron-phonon coupling $\lambda_{epc}$ vs $x$ for (a) Bi2201 and (b) LSCO.  Various phases are 
identified by their dominant ${\bf q}$-vectors, denoted by different symbols: 
diagonal [vertical] ANN phase, with wave vector ${\bf q}$ = $(\delta ,\delta )$ 
[$(\delta ,0)$] as circles [squares]; and diagonal [vertical] near-$(\pi ,\pi )$ phase, ${\bf q}$ = $(\pi -\delta ,\pi -\delta )$ [$(\pi ,\pi -\delta )$] as diamonds [triangles]. Dashed lines in (a) indicate transitions between different symmetries, while dotted line corresponds to $\gamma =3.25$.  (c) Hubbard-index $H$ for Bi2201 (filled red circles) and LSCO (open violet squares) as a function of doping $x$; solid blue line represents 0.1$/x$.
}
\label{fig:8}
\end{figure}

The phase diagram can also be described in terms of a critical electron-phonon coupling $\lambda_{epc}$  vs $x$ for fixed $U$, as in Fig.~\ref{fig:8}.  As doping changes, the threshold $q$-vector varies, and its symmetry can also change, as denoted by the different symbols in the figure.  These phase diagrams show a close resemblance to the magnetic phase diagrams, including the dominant $q$-vectors. However, there are some notable differences.  Whereas $U$ drives magnetic instabilities (in the Stoner denominator), $U$ acts to oppose charge instabilities, which are instead driven by the electron-phonon coupling $\lambda_{ep}$.  The dotted line in Fig.~\ref{fig:8} corresponds to our lower estimate for the experimental $\lambda_{ep}=0.57$; the upper limit, $\lambda_{ep}$=5.4, is off the scale of the figure.  Incorporation of long-range Coulomb interaction might shift the critical value of $\lambda_{ep}$ to larger values, but is unlikely to introduce any qualitative change in the phase diagram.  In comparison to experimental CDW phases, a value of $\lambda_{ep}$ near our lower estimate seems likely.

We briefly comment on the overall doping dependence of the phase diagram.  The minimum value of $\lambda_{epc}$ occurs when the VHS lies at the Fermi energy, $x=x_{VHS}$. Near half filling, a large $U$ leads to strong screening of charge excitations [large $U_{GA}$], which rapidly quenches the e-ph instability.  At $U=U_{BR}$ and $x\rightarrow 0$, $\lambda_{epc}\rightarrow\infty$.  As doping increases, $U_{GA}$ decreases and the CDW nesting effect takes over, so that near $x_{VHS}$, the e-ph instability is nearly $U$-independent.  For $x>x_{VHS}$, there is a residual nesting effect, now largest near $(\pi ,\pi )$, but this falls off rapidly for higher $x$.

\subsection{‘Hubbard index’}

The current results, especially those of Fig. 3, provide insight into the role of $U$ in suppressing transitive charge fluctuations.  The regime $U\ge U_{BR}$ is particularly important as Hartree-Fock calculations have difficulty capturing the underlying physics.  
As hole doping increases, electrons can hop around without causing double occupancy, so that at some point $U$ becomes unimportant in suppressing the CO.  This is not captured in HF, where the Stoner denominator remains $1+U\chi$ at all dopings.  In contrast, the GA solution shows a clear crossover, Fig.~\ref{fig:8}, to a regime where the critical electron-phonon coupling becomes independent of $U$.  Here we quantify this crossover in terms of a `Hubbard index' defined by 
\begin{equation}
H=\frac{\partial\ln{\lambda_c}}{\partial\ln{U}}|_{U=U_{BR}},
\label{eq:30a}
\end{equation}
which measures how sensitive CO is to $U$ at the Brinkman-Rice energy, and it may be approximated as
\begin{equation}
H\sim\frac{U_{BR}}{\lambda_c}\frac{\Delta\lambda_c}{\Delta U}
\label{eq:30b}
\end{equation}
for the two highest $U$s in Fig.~\ref{fig:8}.  While this should be accurate at large $x$, at $x=0$, $\lambda_c$ diverges, so that $\delta\lambda_c/\lambda_c\sim 1$, while $\delta U/U\rightarrow 0$, and $H\rightarrow\infty$ -- that is, the undoped cuprates become incompressible as $U\rightarrow U_{BR}$.  In contrast, our approximate $\Delta U/U=0.25$, so that $H\le 4$.  Bearing that in mind, we show the approximate $H$ in Fig.~\ref{fig:8}(c) as filled circles (Bi2201) or open squares (LSCO).  Remarkably, for both materials, $H\simeq 0.1/x$ (solid line).  This is quite different from the Hartree-Fock expectation, $H\sim constant$, dotted line.  If one looks closely at the data, both curves show hints of a plateau for $0.1\le x\le 0.2$, but for larger $x$, $U$ has an anomalously small effect on CO.

While $U=U_{BR}$ has a well defined meaning only at $x=0$ as a crossover, or as a phase transition in mean-field or infinite dimensions, it still represents a regime of strong suppression of double occupancy $D$. We find that when $U=U_{BR}$, $D$ is approximately 
 \begin{equation}
D=\frac{x}{16(1+x)^2},
\label{eq:30c}
\end{equation}
while the renormalization function $z_0$ is
 \begin{equation}
z_0^2=\frac{x(3+2x)}{(1+x)^2},
\label{eq:30d}
\end{equation}
which is somewhat larger than its $D=0$ value $2x/(1+x)$.


\section{Relation to experiment in high-T$_c$ cuprates}\label{secexp}
\subsection{Charge-Magnetic Crossover in Underdoped Cuprates}\label{seccross}

As noted in Sections \ref{secres1} and \ref{secres2}, for most cuprates there 
should be {\it two} kinds of competing density wave orders, with a transition 
as a function of doping between an incommensurate phase with $q$ near 
$(\pi,\pi)$ (NNN) and an antinodal nesting (ANN) phase.  This is true for
both magnetic instabilities\cite{Gzm1} and CDWs.  The NNN-ANN crossover 
occurs  near $x\sim 0.12-0.16$ for the magnetic phases and at a somewhat lower 
doping for the CDWs.  The question arises, which density wave has lower energy?

At half-filling, the answer is clear, and in good agreement with experiment.  Due to the large Hubbard $U$, the CDW is only marginally stable, whereas $(\pi,\pi)$ AFM order can open a full gap over the FS, leading to a much greater energy lowering.  Thus, near half filling we expect NNN magnetic order for all cuprates.  As doping increases, the pseudogap gets much smaller, and most experiments see a weakening of magnetic fluctuations.  At the same time, with increasing doping, the suppression of CDW order by $U$ becomes relatively less important. Finally, CDW order tends to be commensurate with the lattice, hence belonging to the Ising universality class, which is more robust against fluctuations than magnetism which always breaks an SO(3) continuous symmetry. Given the uncertainty in $\gamma$, it will be hard to determine the exact crossover, but it is likely that the CDW will win out at higher doping.

Recently, there has been considerable experimental evidence for such a transition, between a low-doping SDW phase and a higher-doping ANN CDW phase, both in the Bi-cuprates and in YBCO. 
Here, we will summarize the experimental evidence for such a crossover, while in the remainder of this paper we will explore the consequences of this assumption. In deeply underdoped Bi2212, hints of this transition have been observed in recent scanning tunneling microscopy (STM) studies: Ref.~\cite{ParkYaz} found that the phase we identify here as the ANN CDW seems to weaken below 1/8th doping, while in a similar doping range in CCOC Ref.~\cite{kos12} found that islands of a phase with very weak CO and without $C_4$ symmetry breaking become more prevalent with reduced doping
-- suggestive of domains of predominantly magnetic order, invisible to STM.

Results for YBCO are even more interesting.  A CDW phase has been found\cite{YCDW2,YCDW3,YCDW4,YCDW5,YCDW6,YCDW1} in the doping range near $x=1/8$ where quantum oscillations have been observed\cite{QOsc,QOsc3,QOsc4,Seba1,Seba2,Seba3}.  In the absence of superconductivity, the CDW correlation length $\xi$ remains finite, growing as $T$ decreases\cite{YCDW2}, as expected for a 2D system from the Mermin-Wagner theorem\cite{MW,MK7}. The results are reminiscent of the growth of $(\pi,\pi)$ AFM order in electron-doped cuprates, except that in the latter case a transition to long range magnetic order occurs in underdoped samples, where superconductivity is suppressed \cite{GrevenNCCO}.  In YBCO, when an external magnetic field is used to suppress the superconducting (SC) order, the CDW correlation length is found to increase \cite{YCDW1,YCDW3,YCDW5}.

In this connection, we further note the following points: (a) The nesting vectors are very similar as a function of doping for all cuprates studied, Fig.~\ref{fig:1neu}, down to details of vertical vs diagonal nesting and 1-$q$ vs 2-$q$ order (see Subsection 4.2); (b) The nesting seems to follow the bonding FS of YBCO\cite{YCDW4,YBCOAB}, just as we have found in Bi2212 (Fig.~\ref{fig:1neu}); (c) Ref.~\cite{YCDW6} reports that in the Ortho-II phase in YBCO$_{6.54}$ CDW order along the $b$-axis and SDW order along the $a$-axis are simultaneously present at zero field.  This is similar to the two types of patches observed in extremely underdoped Bi2212\cite{eno12}, which we ascribed to competing CDW and SDW orders; and, (d) In Bi2212, the CDW order is strongest near 1/8th doping, but persists well into the overdoped regime\cite{ParkYaz}.  In YBCO, the CDW effects also peak near $x=1/8$.  The range has been estimated best from measurements of the Hall coefficient, which in YBCO becomes negative at low temperatures in the doping range $\sim0.08\le x\le\sim 0.16$, and this is considered to be due to FS reconstruction associated with CO \cite{HallLif}. At the same time magnetic fluctuations are found in the region below $x=0.08$, associated with the competing near-$(\pi,\pi)$ order.\cite{YBCOmag}

We thus adduce that there is considerable evidence for a crossover as a function 
of doping between a predominantly magnetic phase with incommensurate, 
near-$(\pi,\pi)$ order to an ANN CO phase as doping is increased, 
both in YBCO and in Bi2201/Bi2212, consistent with our nesting model. One should keep in mind, however, that the current experimental situation remains fluid.

\subsection{Nesting Vectors}

For electron doped cuprates, the dominant nonsuperconducting order remains 
commensurate at $(\pi, \pi )$, associated with the magnetic instability, and there has been little evidence for any secondary CO.  The undoped case is special, since the full FS can be gapped and conventional nesting plays no role.  Nevertheless, the optimal energy lowering is associated with $q=(\pi,\pi)$,\cite{Gzm2} which again is the magnetic ordering vector.  Similarly, for hole-doped LSCO the primary order is magnetic, with CO forming on antiphase boundaries at a near-$(\pi,\pi)$ incommensurate $q$-vector\cite{Tranq}.  However, LSCO is a special case, which will be considered  elsewhere\cite{zx9LSCO}, and here we will limit ourselves to the discussion of  CDW modulations in other hole-doped cuprates.

\begin{figure}
\includegraphics[width=10cm,clip=true]{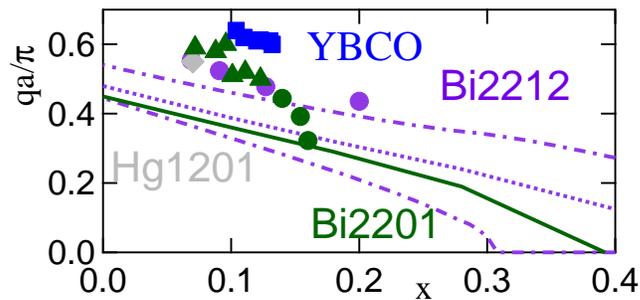}
\caption{(Color online.)
Nesting vectors for ANN $(q,q)$ CO for Bi2201 (green solid line), Bi2212 with (violet dot-dashed) or without (violet dotted) bilayer splitting, compared to experimental $(0,q)$ CO vectors for Bi2201 (green circles\protect\cite{Wise} or triangles\cite{qSilva}), Bi2212 (violet  
circles\protect\cite{JCD}), YBCO (blue squares\cite{YCDW2,YCDW3,YCDW4,YCDW6}), and Hg1201 (silver diamond)\cite{qGrev}.  Band parameters for the theoretical calculations are from Refs.~\protect\cite{ZX01} and ~\protect\cite{Arun3}.  Results for Hg1201 and YBCO are after Ref.\cite{qGrev}.
} 
\label{fig:1neu} 
\end{figure}

In contrast, in Bi2201 and Bi2212, STM studies are sensitive to charge modulations, and we find that the associated $q$-vector is consistent with an important role of ANN nesting in CO. Figure~\ref{fig:1neu} plots the calculated doping dependence of the ANN charge nesting vectors for the Bi cuprates,\cite{ZX01,Arun3} displaying a strong material dependence.  Shown also are the experimental superlattice periodicities for CO found in the tunneling spectra in Bi2201\cite{Wise} (green circles) and Bi2212\cite{JCD} (violet circles), and compared to other measures of the CDW $q$-vector for YBCO (blue squares)\cite{YCDW2,YCDW3,YCDW4,YCDW6}, Bi2201 (green triangles)\cite{qSilva}, and Hg1201 (silver diamond)\cite{qGrev}.  Clearly the experimental superlattices in the Bi-cuprates are close to the predicted ANN periodicities, although in Bi2201 there are hints of shifts to nearby commensurate values. In Bi2212 there are two nesting vectors, associated with bonding and antibonding combinations of the bilayer-split bands, and the experimental data fall close to the bonding band nesting vector.  For both materials, the observed $q$-vector follows the doping dependence of ANN nesting, and is incompatible with an interpretation as a secondary CO as in LSCO, having the wrong doping dependence.

It should be noted that the experimental $q$-vectors represent vertical $(q,0)$ ANN nesting, whereas the theory predicts nesting at $(q',q')$, where $q'$ is close to $q$ in magnitude.  We believe that this is a question of the evolution of the CDW ground state as $U_{eff}$ is increased above threshold, in which case vertical ANN nesting is energetically favored (\ref{appb}).  A similar competition of 1-$q$ vs 3-$q$ nesting has been found in NbSe$_2$\cite{NbSe2}.

A closely related issue for both Bi-cuprates and YBCO is whether the CDWs are modulated along a single $q$ vector, either along the $x$ or the $y$ direction (denoted 1-$q$ nesting),\cite{YCDW6,YCDW1} or whether there is modulation simultaneously along both the $x$- and $y-$ axes (2-$q$ nesting).\cite{YCDW2,YCDW3,YCDW4,YCDW5}  While some experiments cannot distinguish true $2-q$ nesting from patches of $1-q$ ($x$ or $y$) CDWs, others can\cite{YCDW5}. This point is discussed further in Appendix B.

\subsection{Kohn Anomalies and Soft Phonons}

\begin{figure}
\includegraphics[width=9cm,clip=true]{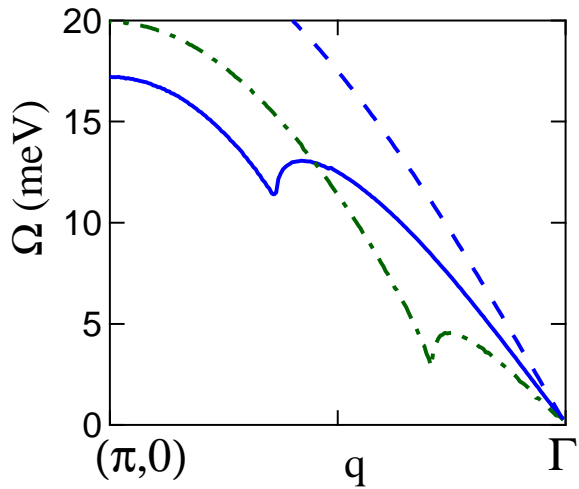}
\caption{(Color online.)
Longitudinal phonon dispersion in LSCO for $U=0.5U_{BR}$, replotted from Fig.~\ref{fig:6}, comparing the bare (blue dashed line) and dressed dispersions for $x$ = 0.4 (solid blue line) and 0.05 (green dotted line).  For intermediate dopings, the anomaly becomes unstable  
}
\label{fig:44}
\end{figure}

To compare the Kohn anomaly to experiment, it must be kept in mind that our model involves effective acoustic modes of a copper plane, which must be embedded into the full phonon band dispersions.  That means the Kohn anomaly will show anticrossing behavior with the different bands.  Here we provide one example of this.  
In Fig.~\ref{fig:44}, we plot an expanded view of the Kohn anomaly in our model. In particular, the anomaly along the $\Gamma\rightarrow (\pi ,0)$ longitudinal branch in LSCO, dark blue curve in Fig.~\ref{fig:6}(a) or Fig.~\ref{fig:6}(d), can be compared with the experimental results of Ref.~\cite{Rez}. While the shape and position are qualitatively correct, some differences can be expected due to our oversimplified phonon model.  In particular, the experimental Kohn anomaly is in an LO branch, whereas our model only has acoustic branches.  Since the bare LO branch has a maximum at $\Gamma$ and decreases towards $(\pi ,0)$, this reverses the left-right asymmetry of the Kohn anomaly.  To see this more clearly, we have replotted the data of Fig.~\ref{fig:6} in Fig.~\ref{fig:44}(b) with a reversed horizontal axis.  A second difference is that the experimental Kohn anomaly softens, but does not go unstable, unlike the theory.  This is again an anticrssing effect, and is discussed further in SOM Section III.

\section{Discussion and Conclusion}\label{secconcl}

\subsection{`Non-BCS' CDWs}
CDWs in cuprates seem anomalous when compared to a BCS-like mean-field picture, which predicts a second order transition with a diverging correlation length, and the ratio of the $T=0$ CDW gap to $T_{CDW}$ of $\eta_{CDW}\equiv 2\Delta_{CDW}(0)/k_BT_{CDW}=3.53$.  In practice, however, such a description hardly ever works even for conventional CDWs in that the correlation length does not diverge and the gap ratio $\eta_{CDW}$ is typically $>>3.53$.  This was originally discussed in 2H-TaSe$_2$ by McMillan,\cite{CDWsc1} who suggested that the short correlation length means that the electron couples to phonons with a wide range of $q$-values, and that the transition is therefore controlled by the phononic and not electronic entropy.  He then modeled the transition as a double transition, first to short-range order at a high $T_{SRO}$ consistent with the BCS ratio, and then to long-range order at a much lower temperature.  The phonon entropy effect is now understood as a breakdown of the RPA due to mode coupling effects.\cite{MCA1,MCA2}  For a 2D system, mode-coupling effects account for the Mermin-Wagner physics, suppressing the transition to $T=0$.\cite{MW,MCA2,MK7}  2D systems are also sensitive to impurity effects, which can further limit correlation lengths.\cite{ImMa}

If the cuprates display a `conventional' CDW instability, we should expect: (a) the transition is characterized by phonon softening; (b) the phonons will soften over a range of $q$-values,\cite{CDWsc3} where the range is related to the electron correlation length--perhaps associated with an order-disorder transition and a central peak\cite{centpk}; (c) the electronic correlation length will not diverge at the transition; and (d) the gap ratio should be anomalously large, with the magnitude of the anomaly also related to the correlation length. In both Bi2212 and YBCO the CDW appears to have an anomalously small correlation length.\cite{zx8a,YCDW4}

\subsection{Purely Electronic CDWs}

We comment on recent papers on purely electronic CDW models\cite{Metlitski,Pepin,LaPlaca,Hayward,Bulut,wang14,allais14}, showing how our work relates to these papers.  One paper\cite{Seamus_cond1} noted, ``because the $Q\ne 0$ modulations exhibit wave vectors generated by scattering regions (`hot spots') moving along the $k$-space lines $(\pm\pi,0)\rightarrow (0,\pm\pi)$, FS nesting provides an inadequate explanation for the cuprate density waves."  Our analysis, however, indicates otherwise in that diagonal hot spot nesting was predicted from a nesting model and its origin carefully described in Ref.~\cite{Gzm1}.  The issues raised in Ref.~\cite{Gzm1} concerning vertical vs diagonal CDWs and one- or two-$q$ order seem to arise in all the current CDW models.  In order to stress that the `hot-spot' nesting is a FS effect unrelated to the $(\pi,\pi)$ magnetic order, a different argument is presented in Appendix B.

All the preceding models are based on an assumed $(\pi,\pi)$-dominated spin susceptibility.  Within a Stoner-type framework, this would suggest the presence of susceptibility peaks at $(\pi,\pi)$ at all dopings, a quantum critical point $x_c$ when $U\chi_{(\pi,\pi)}(T=0)\sim 1$, and strong commensurate fluctuations for $x>x_c$.  However, this is not found to be the case in most cuprates\cite{Gzm1}: as doping increases, there is a crossover to a regime where the `hotspot' susceptibility is the largest, and the near-$(\pi,\pi)$ fluctuations are cut off.  Ironically, only in LSCO is the hotspot susceptibility weak and the near-$(\pi,\pi)$ fluctuations dominate for all dopings.  However, in LSCO the CDW seems to be absent, and the high-doping regime is consistent with a spin-density wave with $q$-vector given by the incommensurate $(\pi,\pi-\delta)$ nesting vector.\cite{Gzm1}

While the above models all describe purely electronic CDWs, the observed CDW couples strongly to phonons.  Indeed, the x-ray diffraction intensities of Ref.~\cite{YCDW4} could only be explained by assuming that the CDW was accompanied by a conventional Peierls distortion, which increased the x-ray intensity by a factor of $\sim$600.  Hence, an improved model should incorporate effects of both the electronic CDW and the accompanying lattice distortion.

\subsection{Conclusions}

We have examined the charge response of the cuprates within the framework of the Hubbard model using the time-dependent Gutzwiller approximation, where effects of a finite electron-phonon coupling are included for the first time. The resulting ANN CDWs provide a good model for the higher-doping regime of the pseudogap phase in most hole doped cuprates.  Specifically, the ANN phase in the Bi cuprates captures the experimentally observed doping dependence of incommensurability, and the predicted FS is found to be consistent with that seen in QO measurements.  A secondary magnetic order (SOM Section I) enhances the resemblance to a conventional stripe phase. Although the CDW is found to have a $d$-wave structure factor\cite{Seamus_cond2} experimentally, this additional modulation is absent in our CDW model, but such a modulation is expected to arise from coupling to a shear strain as discussed in Ref.~\cite{zx8a}.  We note that a shear strain will greatly complicate the cuprate phase diagram by adding, for example, an $s$-wave component to the superconducting gap.\cite{RMV} While the early Peierls calculations found nesting features in the Lindhart susceptibility, these features are reflected more indirectly in the present calculations as they arise from a $q$-dependent phonon softening.  Despite this, the resulting nesting vectors are very similar to those found for magnetic excitations, with only minor shifts due to the $q$-dependence of $U_{eff}$.  Much more dramatic effects of $U_{eff}(q)$ are possible.\cite{Rudi}
Finally, we note that it will be interesting to examine how `nematic' phenomena in the cuprates may be related to the CDWs.

\appendix
\section{Incorporation of phonons in the TDGA formalism}\label{subsecga}
\subsection{Formalism}
Our starting point is the energy functional for the model Eq. (\ref{eq:mod})
evaluated within the Gutzwiller approximation (GA)
\begin{equation}\label{eq:eega}
E^{GA}=E^{Hubbard}+E^{SSH}_{e-ph}+E_{ph}.
\end{equation}
The last term $E_{ph}$ is the phonon part Eq. (\ref{eq:phonly}),
which does not depend on the electronic part of the wavefunction.
We have included this term here in order to account for the renormalization of
the elastic constant due to electron-phonon interactions. Here, 
\begin{equation}
E^{Hubbard}=\sum_{i,j,\sigma}=t_{ij}z_{i,\sigma}z_{j,\sigma}\rho_{i,j,\sigma}
+U\sum_i D_i
\end{equation}
is the Hubbard model contribution where the renormalization factors
\begin{equation}\label{eq:hop}
z_{i\sigma}(\rho, D)=\frac{\sqrt{(\rho_{ii,\sigma}-D_i)(1-\rho_{ii}+D_i)}+
\sqrt{(\rho_{ii,-\sigma}-D_i)D}_i}{\sqrt{\rho_{ii,\sigma}(1-\rho_{ii,\sigma})}},
\end{equation}
depend on both
the density matrix $\rho_{i,j,\sigma}=\langle c_{i\sigma}^\dagger 
c_{j,\sigma}\rangle$ and the double occupancy
variational parameters $D_i$, and   
$\rho_{ii}\equiv\sum_\sigma \rho_{ii,\sigma}$.
For the Su-Schrieffer-Heeger coupling, the GA for 
the electron-phonon interaction of Eq.~(13) 
can be rewritten as
\begin{equation}
E^{SSH}_{e-ph}=-\sum_{i,j,\sigma,\mu}t_{i,j}\gamma_{i,j}{(u_j^{\mu}-u_i^{\mu})\over r_{i,j}} 
z_{i,\sigma}z_{j,\sigma}(\rho_{i,j,\sigma}+\rho_{j,i,\sigma}),
\label{eq:7}
\end{equation}
where $\gamma_{i,j}=\partial \ln{(t_{i,j})}/\partial\ln{(r_{i,j})}<0$ and $\mu =x,y$.

The TDGA involves an expansion of the energy functional
of Eq. (\ref{eq:eega}) up to second order in the density, double occupancy, and
lattice fluctuations:
\begin{equation}\label{eq:eeff}
\delta E^{GA}=E_0 + H^{GA}\delta\rho + \delta E^{(2)}_{Hubbard}
+\delta E^{(2)}_{e-ph}+E_{ph}\,. 
\end{equation}
Here we have defined an effective `Gutzwiller Hamiltonian' 
$H^{GA}=\partial E^{GA}/\partial\rho$
which describes the particle-hole excitations at the level of the
GA. In the following RPA-like treatment, 
the bare (i.e. GA) susceptibilities are evaluated from $H^{GA}$  
whereas interaction effects are contained in the 
second order contributions, which due to translational
invariance will now be evaluated in momentum space.

For the Hubbard term one finds
\begin{eqnarray}
E^{(2)}_{Hubbard}={1\over N}\Bigl[{1\over 2}\sum_qY_q\delta\rho_q\delta\rho_{-q}+z_0z_D'
\sum_q\delta D_q\delta T_{-q}
\nonumber \\
+{1\over 2}z_0(z'+z_{+-}')\sum_qY_q\delta T_q\delta\rho_{-q}
\nonumber \\
+\sum_qL_q\delta\rho_q\delta D_{-q}
+{1\over 2}\sum_qU_q\delta D_q\delta D_{-q}\Bigr],
\label{eq:1}
\end{eqnarray}
where various coefficients are defined in \ref{appelel} and  
the relevant fluctuation modes are the local density fluctuations 
$\delta\rho_q ={1\over N}\sum_{k,\sigma}\delta\rho_{k+q,k,\sigma}$,
the intersite charge fluctuations
$\delta T_q ={1\over 
N}\sum_{k,\sigma}(\epsilon_{k\sigma}^0+\epsilon_{k+q,\sigma}^0)\delta\rho_{k+q,k,\sigma}$,
and the double occupancy fluctuations $\delta D_q$.  We also define 
$\delta\rho_{k+q,k}=\sum_{\sigma}\delta\rho_{k+q,k,\sigma}$ with 
the density matrix $\rho_{kk',\sigma}=\langle c_{k\sigma}^\dagger c_{k'\sigma}\rangle$. The fluctuation contribution for the electron-phonon interaction takes the form
\begin{equation}
\delta E^{(2)}_{SSH}={1\over N}\sum_{q\mu}Q_q^{\mu}[\sum_k
f_{k,k+q,\mu}^{(1)}\delta\rho_{k,k+q}+
f_{q,\mu}^{(2)'}\delta\rho_{-q}+h_{q,\mu}\delta D_{-q}],
\label{eq:8}
\end{equation}
where 
\begin{equation}
f_{k,k+q,\mu}^{(1)}=2iz_0^2f_{k,k+q,\mu}^{(0)},
\label{eq:9}
\end{equation}
\begin{equation}
f_{q,\mu}^{(2)'}=-iz_0(z'+z'_{+-})f_{q,\mu}^{(0)},
\label{eq:10}
\end{equation}
\begin{equation}
h_{q,\mu}=-i2z_0z'_Df_{q,\mu}^{(0)}\,.
\label{eq:11}
\end{equation}
$Q_{q}^{\mu}$ is the Fourier transform of $u_i^{\mu}$, 
$f_{k,k+q,\mu}^{(0)}Q_{q}^{\mu}$ is the Fourier transform of $f_{i,j,\mu}=-t_{i,j}\gamma_{i,j} 
(u_j^{\mu}-u_i^{\mu})/r_{i,j}$, and $f_{q,\mu}^{(0)}Q_q^{\mu}$ is the Fourier 
transform of 
$f_{i,\mu}=\sum_jf_{i,j,\mu}$.  Explicit expressions for $f_{k,k+q,\mu}^{(0)}$ and 
$f_{q,\mu}^{(0)}$ are given in \ref{ap:defph}. Finally, the double occupancy fluctuations can 
be eliminated by an 
antiadiabatic approximation\cite{SeiLo}, where one assumes that the
fluctuations are faster than other degrees of freedom. 
Since we will be concerned with the static limit in the present paper it
is always justified to take this antiadiabatic limit,
\begin{equation}
{\partial E^{(2)}\over\partial D_{-q}}=0,
\label{eq:4}
\end{equation}
which allows us to express the double occupancy via the density fluctuations
\begin{equation}
\delta D_q=-(L_q\delta\rho_q+z_0z_D'\delta T_q+\sum_{\mu}h_{q\mu}Q_q^{\mu})/U_q.
\label{eq:12}
\end{equation}

Inserting Eq. (\ref{eq:12}) into Eq. (\ref{eq:eeff}) yields an
energy functional 
\begin{equation}\label{eq:eeff2}
\delta E^{GA}=E_0 + H^{GA}\delta\rho + \delta \tilde{E}^{(2)}_{el-el}
+\delta \tilde{E}^{(2)}_{e-ph}+\tilde{E}_{ph}
\end{equation}
comprising an effective electron-electron
interaction $\delta \tilde{E}^{(2)}_{el-el}$, a correlation renormalized  
electron-phonon coupling $\delta \tilde{E}^{(2)}_{e-ph}$ and an 
effective phonon part $\tilde{E}_{ph}$. The electron-electron interaction then is:
\begin{equation}                                                             
\delta \tilde{E}^{(2)}_{el-el}= {1\over 2N}\sum_{\bf q}
\Bigl(\begin{array}{c}\delta\rho_{\bf q}\\
      \delta T_{\bf q}
\end{array}\Bigr)^T 
\Bigl(\matrix{A_{\bf q}&B_{\bf q}\cr
                             B_{\bf q}&C_{\bf q}\cr}\Bigr)
\Bigl(\begin{array}{c}\delta\rho_{\bf -q}\\                            
      \delta T_{\bf -q}
\end{array}\Bigr),                                           
\label{eq:A3}                                                                
\end{equation}                                                               
and $A_{\bf q}$, $B_{\bf q}$ and $C_{\bf q}$ are defined in 
\ref{appelel}. Elimination 
of the double occupancy fluctuations also adds a new term to the
electron-phonon coupling [cf. Eq. (\ref{eq:8})]:
\begin{eqnarray}
\delta \tilde{E}^{(2)}_{e-ph}&=&{1\over N}\sum_{q\mu}Q_q^{\mu}[\sum_k
f_{k,k+q,\mu}^{(1)}\delta\rho_{k,k+q}+
f_{q,\mu}^{(2)'}\delta\rho_{-q}+h_{q,\mu}\delta D_{-q}] \nonumber \\
&-&{1\over N}\sum_{q\mu}f_{q\mu}^{(3)}\delta T_qQ_{-q}^{\mu},
\label{eq:14}
\end{eqnarray}
with
\begin{equation}
f_{q\mu}^{(3)}={z_0z'_Dh_{-q\mu}\over U_q}=-2iC_qf_{q,\mu}^{(0)}\,.
\label{eq:15}
\end{equation}
The phonon part $\tilde{E}_{ph}=E_{ph}+\delta E_{ph}$ gets an 
additional contribution which
renormalizes the phonon frequency by
\begin{eqnarray}
\delta E_{ph}=-\sum_{\mu,\nu}h_{q\mu}h_{-q\nu}/U_qQ_{q}^{\mu}Q_{-q}^{\nu}
\nonumber \\
={1\over 2}\sum_{\mu,\nu}\delta K^D_{\mu,\nu,q}Q_{q}^{\mu}Q_{-q}^{\nu},
\label{eq:16}
\end{eqnarray}
with $\delta K^D_{\mu,\nu,q}=C_q\beta_{q\mu}\beta^*_{q\nu}$
and 
\begin{eqnarray}
\beta_{q\mu}=-2if_{q,\mu}^{(0)}.
\label{eq:21}
\end{eqnarray}

The renormalized phonon frequencies become
\begin{eqnarray}
\Omega_{q\pm}^2 = \frac{1}{2}(\Omega_{qx}^{2}+\Omega_{qy}^{2})
\nonumber \\
\pm \frac{1}{2}
\sqrt{(\Omega_{qx}^{2}-\Omega_{qy}^{2})^2+4\delta K_{xy}\delta K_{yx}/M^2} .
\label{eq:27}
\end{eqnarray}
where $\Omega_{q\mu}^{2}=\Omega_{0q\mu}^{2}+\delta K_{\mu\mu}$.
For ${\bf q}$ along $\Gamma\rightarrow (\pi ,0)$ or 
$\Gamma\rightarrow (\pi ,\pi)$, the phonon 
renormalization is purely longitudinal or purely transverse, and the former 
effect is dominant.  
Along $(\pi ,0)\rightarrow (\pi ,\pi)$, the modes mix and the softer transverse 
mode can go unstable first.

Finally, one can define the response functions 
$$
\chi_{ij}(\qvec)=\frac{-i}{N} \int dt \langle {\cal T}
\delta X^i_\qvec(t) \delta X^j_{-\qvec}(0)\rangle,
$$
which are associated with the density fluctuations $\delta X^1_\qvec\equiv
\delta\rho_q$ and $\delta X^1_\qvec\equiv
\delta T_ \qvec$. The bare susceptibilities $\chi^0_{ij}(\qvec)$ are
then evaluated at the GA level (i.e. based on $H_{GA}$), 
whereas the dressed ones can be obtained
following the standard RPA for calculating
the excitations of interacting electrons coupled to phonons
(see e.g. \cite{mahan}).

\subsection{Abbreviations for the electronic interaction kernel}
\label{appelel}

Elements of the interaction kernel in Eq.~\ref{eq:A3}
are given by 
\begin{equation}
A_{\bf q}=Y_{\bf q}-{L_{\bf q}^2\over U_{\bf q}},
\label{eq:G21}
\end{equation}
\begin{equation}
B_{\bf q}={z_0(z'+z'_{+-})\over 2}
-z_0z'_D{L_{\bf q}\over U_{\bf q}},
\label{eq:G22}
\end{equation}
\begin{equation}
C_{\bf q}=-{(z_0z'_D)^2\over U_{\bf q}},
\label{eq:G23}
\end{equation}
where
\begin{equation}
Y_{\bf q}={1\over 2}[(z'+z'_{+-})^2N_{1{\bf 
q}}+z_0(z''_{++}+2z''_{+-}+z''_{--})
N_{2{\bf q}}],
\label{eq:G24}
\end{equation}
\begin{equation}
L_{\bf q}=z'_D(z'+z'_{+-})N_{1{\bf q}}+z_0(z''_{+D}+z''_{-D})
N_{2{\bf q}},
\label{eq:G25}
\end{equation}
\begin{equation}
U_{\bf q}=2z_D^{'2}N_{1{\bf q}}+2z_0z''_{D}N_{2{\bf q}},
\label{eq:G26}
\end{equation}
\begin{eqnarray}
N_{1{\bf q}}={1\over N}\sum_{{\bf k}\sigma}\epsilon^0_{{\bf k+q}\sigma}n_{{\bf 
k}\sigma},
\label{eq:G17}
\end{eqnarray}
and
\begin{equation}
N_{2{\bf q}}={1\over N}\sum_{{\bf k}\sigma}\epsilon^0_{{\bf k}\sigma}n_{{\bf k}\sigma}
=U_{BR}/8.
\label{eq:G16}
\end{equation}

\subsection{Definitions related to electron-phonon interaction parameters}\label{ap:defph}

For each hopping parameter $t$, $t'$, $t''$, $t'''$, we define corresponding $\alpha’s$ as $\alpha$, 
$\alpha'$, $\alpha''$, $\alpha'''$.  Here for compactness, we have introduced $\alpha=-\gamma/r$, where the $\gamma$s are defined in Eq.~(6).  Then
\begin{equation}
f_{k,k+q,\mu}^{(0)}=
F_{k+q,\mu}-F_{k,\mu},
\label{eq:B1}   
\end{equation}
\begin{eqnarray}
F_{k\mu}=2\alpha ts_{k\mu}(a)
\nonumber \\
+4\alpha't's_{\mu}(a)c_{\nu}(a)
+2\alpha''t''s_{k\mu}(2a) 
\nonumber \\
+4\alpha'''t'''[s_{\mu}(2a)c_{\nu}(a)+s_{\mu}(a)c_{\nu}(2a)],
\label{eq:B2}
\end{eqnarray}
where $\mu ,\nu$ can be either $x$ or $y$, with $\nu\ne\mu$.
\begin{eqnarray}
f_{q,\mu}^{(0)}=
\alpha ts_{q\mu}<c_{kx}+c_{ky}>
\nonumber \\
+2\alpha't's_{q\mu}c_{q\nu}<c_{kx}c_{ky}>
+\alpha''t''s_{2q\mu}<c_{2kx}+c_{2ky}>
\nonumber \\
+\alpha'''t'''[s_{q\mu}c_{2q\nu}+s_{2q\mu}c_{q\nu}
]<c_{kx}c_{2ky}+c_{ky}c_{2kx}>,
\label{eq:B3}
\end{eqnarray}
where $<...>$ means an average 
over occupied ${\bf k}$ states.  Note that Eq.~\ref{eq:B3} follows from 
Eq.~\ref{eq:B2} by averaging over ${\bf k}$, and noting that $<s_{k\mu}>=0$ [so $<F_{k,\mu}>=0$] and $<c_{kx}> =<c_{ky}>$.

\section{Nesting maps and origins of 2-$q$ nesting}\label{appb}

In Section 4.2, we noted that the experimental and theoretical $q$-vectors for the ANN phase point in different directions.
 Here we show how this is related to the question of 1-$q$ vs 2-$q$ nesting. When the hopping $t'$ is large, the paramagnetic cuprates display long nearly parallel regions of FS across $(\pi,0)$ in the antinodal region.  The vertical $(q,0)$ ANN nesting vector takes full advantage of this nesting to produce a nearly 1D Peierls CDW.  
However, this leaves the equivalent FSs near $(0,\pi)$ completely unnested.  In contrast, the theory finds that diagonal nesting at $(q',q')$, where $q'$ is close to $q$ in magnitude, is the optimal single-$Q$ nesting vector, since it allows nesting simultaneously near both AN regions along the $x$ and $y$ axes. This is true at threshold, $\lambda_{ep}=\lambda_{epc}$, where the Stoner criterion allows only single-$q$ nesting. However, the cuprates are already deep in the ordered phase $\lambda_{ep}>\lambda_{epc}$, and there can be a crossover with increasing $\lambda$ from 1-$q$ to 2-$q$ nesting, which we explore in the following. 

Here, the Stoner criterion is of limited value in predicting the dominant instability since it corresponds to the Gaussian level of a corresponding Landau approach. Far away from the ordering transition, higher order terms in the Landau functional become relevant \cite{McMillan}, and may shift the dominant $q$-vector, or even favor competing instabilities with different direction and dimensionality of the CO modulation. For instance, in the magnetic case when $t'/t=-0.2$, with increasing $U$ there is a crossover in the SDW ordering wave vector  from  $(\pi,\pi-\delta)$ to $(\pi-\delta,\pi-\delta)$.\cite{ArrSt}  In the present case, a transition to $(q,0)+(0,q)$ would be advantageous, since it could nest the two antinodal regions much better than single-$q$ nesting at $(q',q')$.  Unfortunately, extending these calculations for CO would require extensive unrestricted HF or GA modeling.  Here we introduce a simple model of the nesting, which nevertheless provides an explanation of why $2-q$ nesting would dominate.

Ref.\cite{Gzm1} shows that a map of the susceptibility is dominated by ridges, which represent a doubled, folded map of the FS, $q=2k_F$.  We briefly discuss the origin of this map and then use it to compare diagonal vs 2-$q$ nesting.  For simplicity we will analyze the Lindhart susceptibility $\chi_0$ at $T,\omega=0$,
$$\chi_0(q)=\sum_k\frac{f(\epsilon_k)-f(\epsilon_{k+q})}{\epsilon_{k+q}-\epsilon_k},$$
where at $T=0$ the Fermi function $f$ becomes a step function at $E_F$.
The numerator is zero unless $\epsilon_k$ and $\epsilon_{k+q}$ are on opposite sides of $E_F$.  The surface contribution arises when $\epsilon_k$ and $\epsilon_{k+q}$ both approach $E_F$, in which case the corresponding denominator in $\chi_0$ vanishes.  


Now we fix $k$ on the FS, so that $\epsilon_k=E_F$.  For an arbitrary direction of $q$, $\epsilon_{k+q}$ will cross the FS again for some magnitude of $q$, unless, e.g., $q$ is tangent to the FS.  However, for a generic point, a small change in $k$ along the FS will lead to a change in the magnitude of $q$, and hence a negligible contribution to the integral over $k$.  Only in special circumstances will the contribution be finite -- for instance, when $q$ is perpendicular to the FS and $q=2k_F$.  This is the conventional nesting, leading to a contribution to the susceptibility inversely proportional to the local FS curvature, or loosely speaking, to the length $L$ of FS that nests in a 2D case.  

Let us apply this to a very simple model of AN nesting.  Let the antinodal part of the FS be flat over a length $L$ in $k$-space, with the two flat sections separated by $q_{AN}$.  If one FS is shifted vertically with respect to the other by $q_{AN}$, the two FSs will nest over a length $L$, so that $\chi_0(q_{AN},0)\sim L$.  If the FS is shifted diagonally, the surfaces will nest over only $L-q_{AN}$, but {\it both the x- and y- antinodal regions will be nested} (double nesting), yielding $\chi_0(q_{AN},q_{AN})\sim 2(L-q_{AN})$.  Hence, for $L>2q_{AN}$, diagonal nesting becomes unstable first.  On the other hand, if there are CDWs along both x and y, we would nest substantially more FS $\chi_0\sim 2L$, and hence 2-$q$ nesting would dominate.  In reality, FS sections are almost never exactly parallel, so diagonal nesting wins at threshold.  However, when $U_{eff}$ is larger than the threshold value $U_c$, more of the FS will be gapped, so the above agrument holds for `nearly-nesting' segments, and 2-$q$ should ultimately win out.

Next, we give an additional argument that diagonal AN nesting is a form of `hot-spot' nesting, but it has its origin in a band structure effect {\it completely unrelated to any underlying $(\pi,\pi)$ AF order}.\cite{Gzm1}  Because the FS has a mirror symmetry about the $\Gamma\rightarrow (\pi,\pi)$ line of the Brillouin zone, the folded FS maps $q=2k_F$ $\rightarrow$ $(2\pi-q_x,q_y)$ or $(q_x,2\pi-q_y)$ will overlap only when $q_x=q_y$ along the zone diagonal.  But when unfolded into the doubled zone, the diagonal becomes the lines $(2\pi,0)\rightarrow (\pi,\pi)$ and $(\pi,\pi)\rightarrow (0,2\pi)$, which are the $q=2k_F$ image of the AF zone boundary.

Finally, we would like to clarify a point of terminology, which is often
confusing in the literature.  For instance, the `packed golf ball' motif shown in the inset of Fig.~1(c) of Ref.~\cite{Wise} is characterized in that paper as
‘checkerboard’ order.  However, a checkerboard implies a density modulation (high-low-high-low) along the Cu-O bond direction. This kind of CO would induce peaks in the Fourier transformed spectra along the diagonals
of the Brillouin zone, whereas the experimental data\cite{Wise}
clearly have the maxima lying along the reciprocal Cu-O bond direction.
The observed pattern is in fact more properly termed `crossed stripes',
having three different charge densities: high in the regions where (charge) stripes cross, low on the sites which are not occupied by the (charge) stripes, and intermediate on other sites where only one stripe is occupied.  For crossed 6$\times$6 stripes (SOM Section 1) this is exactly the 2D pattern observed in Ref.~\cite{Wise}, which also has the dominant Fourier peaks along the reciprocal Cu-O bond direction.

{\bf Acknowledgments} We thank Jenny Hoffman and Y. Kohsaka for many stimulating comments. J. L. acknowledges hospitality by the Aspen Center for Physics under the National Science Foundation’s Grant No. PHYS-1066293.  The work at Northeastern University is supported by the US Department of Energy, Office of Science, Basic Energy Sciences grant number DE-FG02-07ER46352, and benefited from Northeastern University's Advanced Scientific Computation Center (ASCC), theory support at the Advanced Light Source, Berkeley, and the allocation of supercomputer time at NERSC through grant number DE-AC02-05CH11231. J.L. is supported by IIT-Seed project NEWDFESCM, while GS' work is supported by   
the Vigoni Program 2007-2008 of the Ateneo Italo-Tedesco
Deutsch-Italienisches Hochschulzentrum.


\begin{thebibliography}{99}
\bibitem{Tranq}J.M. Tranquada, 
B.J. Sternlieb, J.D. Axe, Y. Nakamura, and S. Uchida, 
Nature {\bf 375}, 561 (1995); 
J.M. Tranquada, J.D. Axe, N. Ichikawa, A.R. Moodenbaugh, Y. Nakamura, and S. Uchida, 
Phys. Rev. Lett {\bf 78}, 338 (1997).
\bibitem{YCDW2}G. Ghiringhelli, M. Le Tacon, M. Minola, S. Blanco-Canosa, C. Mazzoli, N. B. Brookes, G.M. De Luca, A. Frano, D.G. Hawthorn, F. He, T. Loew, M. Moretti Sala, D.C. Peets, M. Salluzzo, E. Schierle, R. Sutarto, G.A. Sawatzky, 
E. Weschke, B. Keimer, and L. Braicovich, Science {\bf 337}, 821 (2012). 
\bibitem{YCDW3}A.J. Achkar, R. Sutarto, X. Mao, F. He, A. Frano, S. Blanco-Canosa, M. Le Tacon, G. Ghiringhelli, L. Braicovich, M. Minola, M. Moretti Sala, C. Mazzoli, Ruixing Liang, D.A. Bonn, W.N. Hardy, B. Keimer, G.A. Sawatzky, and D.G. Hawthorn, Phys. Rev. Lett. {\bf 109}, 167001 (2012).
\bibitem{YCDW4}J. Chang, E. Blackburn, A.T. Holmes, N.B. Christensen, J. Larsen, J. Mesot, Ruixing Liang, D.A. Bonn, W.N. Hardy, A. Watenphul, M. v. Zimmermann, E.M. Forgan, and S.M. Hayden, Nature Physics {\bf 8}, 871 (2012).
\bibitem{YCDW5}D. LeBoeuf, S. Kr\"amer, W.N. Hardy, Ruixing Liang, D.A. Bonn, and C. Proust, Nature Physics {\bf 9},79 (2013), 
\bibitem{YCDW6}E. Blackburn, J. Chang, M. Hucker, A.T. Holmes, N.B. Christensen, Ruixing Liang, D.A. Bonn, W N. Hardy, M. v. Zimmermann, E.M. Forgan, and S.M. Hayden, Phys. Rev. Lett. {\bf 110}, 137004 (2013). 
\bibitem{comin} R. Comin, A. Frano, M. M. Yee, Y. Yoshida, H. Eisaki,
                E. Schierle, E. Weschke, R. Sutarto, F. He, A. Soumyanarayanan,
                Yang He, M. Le Tacon, I. S. Elfimov, Jennifer E. Hoffman,
                G. A. Sawatzky, B. Keimer, A. Damascelli, Science {\bf 343}, 
                390 (2014).
\bibitem{LoSe}J. Zaanen and O. Gunnarsson, Phys. Rev. B 40, R7391
(1989); K. Machida, Physica (Amsterdam) 158C, 192
(1989); H. J. Schulz, Phys. Rev. Lett. 64, 1445 (1990);
D. Poilblanc and T. M. Rice, Phys. Rev. B 39, R9749
(1989); J. Lorenzana and G. Seibold, Phys. Rev. Lett. {\bf 89}, 136401 (2002).
\bibitem{wang14} Y. Wang and A. V. Chubukov, arXiv:1401.0712.
\bibitem{allais14} A. Allais, J. Bauer, and S. Sachdev,
  arXiv:1402.4807,6311.
\bibitem{cas95}C. Castellani, C. Di Castro, and
  M. Grilli. Phys. Rev. Lett. {\bf 75}, 4650 (1995). 
\bibitem{AIP}T. Das, R.S. Markiewicz, and A. Bansil, arXiv:1407.5722, to be published, Adv. in Phys.

\bibitem{correl1}A. Comanac, L. d. Medici, M. Capone, and A. J. Millis, Nat. Phys. {\bf 4}, 287 (2008).
\bibitem{Gzm2}
R.S. Markiewicz, J. Lorenzana, and G. Seibold, Phys. Rev. B{\bf 81}, 014510 (2010).
\bibitem{tanmoyop}T. Das, R. S. Markiewicz, and A. Bansil,
{Phys. Rev. B} {\bf 81}, 174504 (2010).
\bibitem{ASWT}R.S. Markiewicz, Tanmoy Das, and A. Bansil, Phys. Rev. B{\bf 82}, 224501 (2010).
\bibitem{correl2} J. M. P. Carmelo, M. A. N. Ara\'ujo, S. R. White,  
and M. J. Sampaio, Phys. Rev. B {\bf 86}, 064520 (2012).
\bibitem{correl3} G. Seibold, M. Grilli, and J. Lorenzana, Physica C 
                  {\bf 481}, 132 (2012).
\bibitem{br}W.F. Brinkman and T.M. Rice, Phys. Rev. B{\bf 2}, 4302 (1970).
\bibitem{footRef}The GA+RPA approach is not ideally suited for studying `Mott' or spin-liquid phases.  However, a recent beyond-Gutzwiller variational calculation\cite{TBPS} found that, in the $t-t'-U$ Hubbard model at half filling, a spin liquid phase is found only for $U>U_{BR}$ and for $t'<-0.5t$, both values lying outside the cuprate parameter range.  For $-0.5 <t'/t <0$ and any $U$, the only insulating phase found was $(\pi,\pi)$-AFM, turning on near $U\sim 4t$.  For this phase, the GA+RPA phase boundary is in excellent agreement with the beyond-Gutzwiller variational results.\cite{Gzm2}
\bibitem{TBPS}L.F. Tocchio, F. Becca, A. Parola, and S. Sorella, Phys. Rev. B 78 (2008), p. 041101(R) ; 
F. Becca, L.F. Tocchio, and S. Sorella, Proc. HFM2008 Conf., arXiv:0810.0665.
\bibitem{Gzm1}
R.S. Markiewicz, J. Lorenzana, G. Seibold, and A. Bansil, Phys. Rev. B{\bf 81}, 014509 (2010).
\bibitem{SSH}W.P. Su, J.R. Schrieffer, and A.J. Heeger, Phys. Rev. B{\bf 22}, 2099 (1980).
\bibitem{odlsg}E. von Oelsen, A. Di Ciolo, J. Lorenzana, G. Seibold, and M. Grilli, Phys. Rev. B {\bf 81}, 155116 (2010).
\bibitem{Arun3}R.S. Markiewicz, S. Sahrakorpi, M. Lindroos, Hsin Lin, and A. Bansil Phys. Rev. B{\bf 72}, 054519 (2005).
\bibitem{AB1a} A. Bansil, Phys. Rev. Lett.  {\bf 41}, 1670(1978); L. Schwartz and A. Bansil, Phys. Rev. B {\bf 10}, 3261 (1974); R. Prasad and A. Bansil, Phys. Rev. B {\bf 21}, 496 (1980).
\bibitem{AB1b} H. Lin, S. Sahrakorpi, R.S. Markiewicz, and A. Bansil, Phys. Rev. Lett. {\bf 96}, 097001 (2006); S.N. Khanna, A.K. Ibrahim, S.W. McKnight, and A. Bansil, Solid State Commun. {\bf 55}, 223 (1985); L. Huisman, D. Nicholson, L. Schwartz and A. Bansil, Phys. Rev. B {\bf 24}, 1824 (1981).
\bibitem{DLGS}A. Di Ciolo, J. Lorenzana, M. Grilli, and G. Seibold, Phys. Rev. B{\bf 79}, 085101 (2009). 
\bibitem{SeiLo}G. Seibold and J. Lorenzana, Phys. Rev. Lett. {\bf 86}, 2605 (2001).
\bibitem{footlep}The factor of 4 is for ease in comparison with results on transition-metal compounds.\cite{MCA1,MCA2}
\bibitem{MCA1}K. Motizuki and N. Suzuki, {\it Structural Phase Transitions in Layered Transition-Metal Compounds} (Reidel, Dordrecht, 1986).
\bibitem{MCA2}H. Yoshiyama, Y. Takaoka, N. Suzuki, and K. Motizuki, J. Phys. C {\bf 19}, 5591 (1986).
\bibitem{footMcM}Note that this has the McMillan form\cite{McM} of $\lambda_{ep}$, with $g^2\rightarrow <I^2>$, and $K\rightarrow M<\omega^2>$, where $I$ is the electron-phonon interaction, and the brackets denote an appropriate average for superconductivity, whereas our expression is for a single mode.
\bibitem{McM}W.L. McMillan, Phys. Rev. {\bf 167}, 331 (1968).
\bibitem{SeGL}T. P. Devereaux, T. Cuk, Z.-X. Shen, and N. Nagaosa, Phys. Rev. Lett. {\bf 93}, 117004 (2004).
\bibitem{gml11} G. Seibold, M. Grilli, and J. Lorenzana
 Phys. Rev. B{\bf 83}, 174522 (2011)
\bibitem{WYK}C.-Z. Wang, R. Yu, and H. Krakauer, Phys. Rev. B{\bf 59}, 9278  
(1999).
\bibitem{Alig}A.A. Aligia, M. Kuli\'c, V. Zlatic, and K.H. Bennemann, Sol. St.
Commun. {\bf 65}, 501 (1988).
\bibitem{NKF}H. Fukuyama, H. Kohno, B.Normand, and T. Ta\-na\-moto, J. Low-T. Phys. {\bf 99}, 429 (1995);
B.Normand, H. Kohno, and H. Fukuyama, {\it ibid.}, p. 531, and Phys. Rev. B{\bf 53}, 856 (1996).
\bibitem{OKA2}O.K. Andersen, A.I. Liechtenstein, O. Jepsen, and F. Paulsen,
J. Phys. Chem. Solids {\bf 56}, 1573 (1995).
\bibitem{Giustino}F. Giustino, M. L. Cohen, and S. G. Louie,
{Nature} {\bf 452}, 965 (2008).
\bibitem{Lep1}S.-Y. Savrasov,  and O.-K. Andersen,  
Phys. Rev. Lett. {\bf 77}, 4430 (1996).
\bibitem{Lep2}K.-P. Bohnen, R. Heid, and M. Krauss, 
Europhys. Lett. {\bf 64}, 104 (2003).
\bibitem{Lep3}R. Heid, K.-P. Bohnen, R. Zeyher, and D. Manske, 
Phys. Rev. Lett.{\bf 100}, 137001 (2008). 
\bibitem{RG1}H.C. Fu, C. Honerkamp, and D.-H. Lee, EPL (Europhysics Letters) {\bf 75}, 146 (2006). 
\bibitem{trans1}M.L. Kuli\'c and R. Zeyher, Phys. Rev. B{\bf 49}, 4395 (1994);
R. Zeyher and M.L. Kuli\'c, Phys. Rev. B{\bf 53}, 2850 (1996);
G. Seibold, F. Becca, F. Bucci, C. Castellani, C. Di Castro, and M. Grilli, Eur. Phys. J. B {\bf 13}, 87 (2000);
and Ref.~\cite{DLGS}.
\bibitem{QMC1}
M. Grilli and C. Castellani, Phys. Rev. B{\bf 50}, 16880 (1994);
Z.B. Huang, W. Hanke, E. Arrigoni, and D.J. Scalapino, Phys. Rev. B{\bf 68}, 220507 (2003).
\bibitem{ParkYaz}C.V. Parker, P. Aynajian, E.H. da Silva Neto, A. Pushp, S. Ono, J. Wen, Z. Xu, G. Gu, and A. Yazdani, Nature {\bf 468}, 677 (2010). 
\bibitem{kos12}Y. Kohsaka, T. Hanaguri, M. Azuma, M. Takano, J. C. Davis, and H. Takagi, Nature Physics {\bf 8},534 (2012).
\bibitem{YCDW1}T. Wu, H. Mayaffre, S. Kr\"amer, M. Horvati\'c, C. Berthier, W.N. Hardy, Ruixing Liang, D.A. Bonn, and 
M.-H. Julien, Nature {\bf 477}, 191 (2011).
\bibitem{QOsc}N. Doiron-Leyraud, C. Proust, D. LeBoeuf, J. Levallois, J.-B. Bonnemaison, R. Liang, D.A. Bonn, W.N. Hardy, and L. Taillefer, Nature (London) {\bf 447}, 565 (2007).
\bibitem{QOsc3}E.A. Yelland, 
J. Singleton, C.H. Mielke, N. Harrison, F.F. Balakirev, B. Dabrowski, and J.R. Cooper, 
Phys. Rev. Lett. {\bf 100}, 047003 (2008).
\bibitem{QOsc4}A.F. Bangura, J.D. Fletcher, A. Carrington, J. Levallois, M. Nardone, B. Vignolle, P.J. Heard, N. 
Doiron-Leyraud, D. LeBoeuf, L. Taillefer, S. Adachi, C. Proust, and N.E. Hussey, 
Phys. Rev. Lett. {\bf 100}, 047004 (2008).
\bibitem{Seba1}S.E. Sebastian, N. Harrison, C.H. Mielke, Ruixing Liang, D.A. Bonn, W.N. Hardy, and G.G. Lonzarich,
Phys. Rev. Lett. {\bf 103}, 256405 (2009); S.E. Sebastian, N. Harrison, M.M. Altarawneh, C.H. Mielke, Ruixing Liang, D.A. Bonn, W.N. Hardy, and G.G. Lonzarich, PNAS {\bf 107}, 6175 (2010); S.E. Sebastian, N. Harrison, and G.G. Lonzarich, Rep. Prog. Phys. {\bf 75} 102501 (2012); S.E. Sebastian, N. Harrison, Ruixing Liang, D.A. Bonn, W.N. Hardy, C.H. Mielke, and G.G. Lonzarich,  Phys. Rev. Lett. {\bf 108}, 196403 (2012). 
\bibitem{Seba2}S.E. Sebastian, N. Harrison, M.M. Altarawneh, R. Liang, D.A. Bonn, W.N. Hardy, and 
G.G. Lonzarich, Nature Communications {\bf 2}, 471 (2011). 
\bibitem{Seba3}N. Harrison and S.E. Sebastian, Phys. Rev. Lett. {\bf 106}, 226402 (2011).
\bibitem{MW}N. D. Mermin and H. Wagner, Phys. Rev. Lett. {\bf 17}, 1133 (1966).
\bibitem{MK7}R.S. Markiewicz, Phys. Rev. B{\bf 70}, 174518 (2004).  
\bibitem{GrevenNCCO}E.M. Motoyama, G. Yu, I.M. Vishik, O.P. Vajk, P.K. Mang, and M. Greven, Nature {\bf 445}, 186 (2007).
\bibitem{YBCOAB} J. C. Campuzano, L. C. Smedskjaer, R. Benedek, G. Jennings, and A. Bansil, Phys. Rev. B {\bf 43}, 2788(1991).
\bibitem{eno12}M. Enoki, M. Fujita, T. Nishizaki, S. Iikubo, D. K. Singh, S. Chang, J. M. Tranquada, K. Yamada, Phys. Rev. Lett. {\bf 110}, 017004, (2013).
\bibitem{HallLif}D. LeBoeuf, N. Doiron-Leyraud, B. Vignolle, M. Sutherland, B.J. Ramshaw, J. Levallois, R. Daou, F. Lalibert\'e, O. Cyr-Choini\`ere, J. Chang, Y.J. Jo, L. Balicas, Ruixing Liang, D.A. Bonn, W.N. Hardy, C. Proust, and L. Taillefer,
 Phys. Rev. B{\bf 83}, 054506 (2011).
\bibitem{YBCOmag}D. Haug, V. Hinkov, Y. Sidis, P. Bourges, N.B. Christensen, A. Ivanov, T. Keller, C.T. Lin, and B. Keimer, New J. Phys. {\bf 12}, 105006 (2010).
\bibitem{zx9LSCO}R.S. Markiewicz and A. Bansil, unpublished.
\bibitem{ZX01}M. Hashimoto, 
T. Yoshida, H. Yagi, M. Takizawa, A. Fujimori, M. Kubota, K. Ono, K. Tanaka, D.H. Lu, Z.-X. 
Shen, S. Ono, and Yoichi Ando, 
Phys. Rev. B{\bf 77}, 094516 (2008).
\bibitem{Wise}W.D. Wise, M.C. Boyer, K. Chatterjee, T. Kondo, T. Takeuchi, H. Ikuta, Y. Wang, 
and E.W. Hudson, Nature Physics {\bf 4}, 696 (2008).

\bibitem{JCD}Y. Kohsaka, C. Taylor, P. Wahl, A. Schmidt, J. Lee, K. Fujita, J. Alldredge, J. 
Lee, K. McElroy, H. Eisaki, S. Uchida, D.-H. Lee, and J.C. Davis,
Nature {\bf 454}, 1072 (2008).
\bibitem{qSilva}E.H. da Silva Neto, P. Aynajian, A. Frano, R. Comin, E. Schierle, E. Weschke, A. Gyenis, J. Wen, J. Schneeloch, Z. Xu, S. Ono, G. Gu, M. Le Tacon, and A. Yazdani,  Science {\bf 343}, 393 (2014).
\bibitem{qGrev}W. Tabis, Y. Li, M. Le Tacon, L. Braicovich, A. Kreyssig, M. Minola, G. Dellea, E. Weschke, M.J. Veit, A.I. Goldman, T. Schmitt, G. Ghiringhelli, N. Bari{\v s}i'c, M.K. Chan, C.J. Dorow, G. Yu, X. Zhao, B. Keimer, and M. Greven, arXiv:1404.7658.
\bibitem{NbSe2}A. Soumyanarayanan, M.M. Yee, Y. He, J. van Wezel, D.J. Rahn, K. Rossnagel, E.W. Hudson, M.R. Norman, and J.E. Hoffman, Proc. Natl. Acad. Sci. {\bf 110}, 1623 (2013). 
\bibitem{Rez}D. Reznik, L. Pintschovius, M. Ito, S. Iikubo, M. Sato, H. Goka, M. Fujita, K. Yamada, G.D. Gu, and J.M. Tranquada, Nature (London) {\bf 440}, 1170 (2006). 
\bibitem{CDWsc1}W.L. McMillan, Phys. Rev. B{\bf 16}, 643 (1977).
\bibitem{ImMa}Y. Imry and S.-K. Ma, Phys. Rev. Lett. {\bf 35}, 1399 (1975); L.J. Sham and B.R. Patton, Phys. Rev. B{\bf 13}, 3151 (1976).
\bibitem{CDWsc3}F. Weber, S. Rosenkranz, J.-P. Castellan, R. Osborn, R. Hott, R. Heid, K.-P. Bohnen, T. Egami, A. H. Said, and D. Reznik, Phys. Rev. Lett. {\bf 107}, 107403 (2011); M. Leroux, M. Le Tacon, M. Calandra, L. Cario, M-A. M\'easson, P. Diener, E. Borrissenko, A. Bosak, and P. Rodi`ere, Phys. Rev. B{\bf 86}, 155125 (2012).
\bibitem{centpk}M Sato, H Fujishita, S Sato and S Hoshino, J. Phys. C: Solid State Phys. {\bf 18}, 2603 (1985). 
\bibitem{zx8a}R.S. Markiewicz, J. Lorenzana, G. Seibold, and A. Bansil, arXiv:1207.5715.


%














\bibitem{Metlitski}M.A. Metlitski and S. Sachdev, Phys. Rev. B{\bf 82}, 075128 (2010).
\bibitem{Pepin}K.B. Efetov, H. Meier, and C. P\'epin, Nature Phys. {\bf 9}, 442 (2013);
H. Meier, C. P\'epin, M. Einenkel, and K.B. Efetov, arxiv:1312.2010.
\bibitem{LaPlaca}R. La Placa and S. Sachdev, Phys. Rev. Lett. {\bf 111}, 027202 (2013).
\bibitem{Hayward}L.E. Hayward, D.G. Hawthorn, R.G. Melko, and S. Sachdev, arxiv:1309.6639.
\bibitem{Bulut}S. Bulut, W.A. Atkinson, and A.P. Kampf, Phys. Rev. B{\bf 88}, 155132 (2013).
\bibitem{Seamus_cond1}K. Fujita, C.K. Kim, I. Lee, J. Lee, M.H. Hamidian, I.A. Firmo, S. Mukhopadhyay, H. Eisaki, S. Uchida, M.J. Lawler, E.-A. Kim, and J C. Davis,  Science {\bf 344}, 612 (2014).  
\bibitem{Seamus_cond2}K. Fujita, M.H. Hamidian, S.D. Edkins, C.K. Kim, Y. Kohsaka, M. Azuma, M. Takano, H. Takagi, H. Eisaki, S. Uchida, A. Allais, M.J. Lawler, E.-A. Kim, S. Sachdev, and J.C. S\'eamus Davis, arXiv:1404.0362.
\bibitem{RMV}R.S. Markiewicz and M.T. Vaughn, Phys. Rev. B{\bf 57}, R14052 (1998).
\bibitem{Rudi}H.-M. Eiter, M. Lavagnini, R. Hackl, E.A. Nowadnick, A.F. Kemper, T.P. Devereaux, J.-H. Chu, J.G. Analytis, I.R. Fischer, and L. Degiorgi, Proc. Nat. Acad. Sci. {\bf 110}, 64 (2013).
\bibitem{mahan} G. D. Mahan, {\it Many-Particle Physics}, Plenum Press, New York (1981).
\bibitem{McMillan}W.L. McMillan, Phys. Rev. B{\bf 12}, 1187 (1975). 
\bibitem{ArrSt}E. Arrigoni and G.C. Strinati, Phys. Rev. B{\bf 44}, 7455 (1991). 
\end{thebibliography}
\end{document}